\documentclass[twoside]{article}
\usepackage{qic}

\begin{document}
\setlength{\textheight}{8.0truein}    

\runninghead{Semi-loss-tolerant strong coin flipping protocol using EPR pairs}
            {J.J. Ma, F.Z. Guo, Q. Yang, Y.B. Li, Q.Y. Wen}

\normalsize\textlineskip
\thispagestyle{empty}
\setcounter{page}{1}

\copyrightheading{0}{0}{2003}{000--000}

\vspace*{0.88truein}

\alphfootnote

\fpage{1}

\centerline{\bf
SEMI-LOSS-TOLERANT STRONG COIN FLIPPING PROTOCOL USING EPR PAIRS}
\vspace*{0.37truein}

\centerline{\footnotesize
 Jia-Jun Ma$^{1,2}$, {Fen-Zhuo Guo$^{1,2}$}\footnote{gfenzhuo@bupt.edu.cn} , Qian Yang$^1$, Yan-Bing Li$^1$, Qiao-Yan Wen$^1$}
\vspace*{0.015truein}
\centerline{\footnotesize\it 1.State Key Laboratory of Networking and Switching Technology}
\baselineskip=10pt
\centerline{\footnotesize\it Beijing University of Posts and Telecommunications, Beijing, 100876, China}
\baselineskip=10pt
\centerline{\footnotesize\it 2.School of Science, Beijing University of Posts and Telecommunications}
\baselineskip=10pt
\centerline{\footnotesize\it Beijing, 100876, China}
\vspace*{10pt}
\vspace*{0.225truein}
\publisher{(received date)}{(revised date)}

\vspace*{0.21truein}

\abstracts{
\indent In this paper, we present a quantum strong coin flipping protocol. In this protocol, an EPR pair and a quantum memory storage are made use of, and losses in the quantum communication channel and quantum memory storage are all analyzed. We obtain the bias in the fair scenario as a function of $p$, where $p$ is the probability that the particle in Bob's quantum memory storage is lost, which means our bias varies as the degree of losses in the quantum memory storage changes. Therefore we call our protocol semi-loss-tolerant. We also show that the bias decreases with decreasing $p$. When $p$ approaches $0$, the bias approaches 0.3536, which is less than that of all the previous loss-tolerant protocols. Details of both parties' optimal cheating strategies are also given and analyzed. What's more, experimental feasibility is discussed and demonstrated.
}{}{}

\vspace*{10pt}

\vspace*{3pt}
\communicate{to be filled by the Editorial}

\vspace*{1pt}\textlineskip    
\section{Introduction}
\indent
Coin flipping(CF) is a cryptographic primitive which was firstly introduced by Blum in 1981\cite{1}. Its goal is to enable two mutually distrustful and spatially separated parties, usually referred as Alice and Bob, to generate a random bit whose value cannot be controlled by any one of them. That is to say, if both parties are honest, the generated bit must be 0 or 1 with the same probability $\frac{1}{2}$, while even if one party is dishonest, it is guaranteed that the outcome cannot be biased to 0 or 1 with probability 1 by the cheater. Strong CF(SCF)\cite{2,3,4,01,5}, the most common form of CF, requires that a dishonest party, denoted by X, can by no means improve the probability of any value of the bit to be greater than $P_X=\frac{1}{2}+\epsilon_X$. While in a weaker form, which is called weak CF(WCF) \cite{61,6,7}, both Alice and Bob have their preferred outcomes which are opposite and known to each other, and it is required that a dishonest X cannot improve the probability of his or her preferred outcome to be greater than $P_X=\frac{1}{2}+\epsilon_X$. The parameter $\epsilon_X$, which is called the bias of X, quantifies the security of a CF protocol and it must be strictly less than $\frac{1}{2}$, in which case a cheater cannot totally control the outcome. The less $\epsilon=max(\epsilon_A,\epsilon_B)$ is, the securer the protocol is. When we say a CF protocol is fair, we mean the biases for both parties are equal, i.e. $\epsilon_A=\epsilon_B$. A CF protocol is said to be perfect iff $\epsilon_A=\epsilon_B=0$.\\
\indent
	Even though there are many classical approaches dealing with coin flipping tasks, their security is all under the assumption of the complexity of a computational task(many of which may be efficiently solved by a quantum computer in the future). Given unlimited computational power, a cheater can always bias the probability of any outcome to 1, so unconditional secure coin flipping cannot be realized by classical means.\\
\indent
	In the quantum settings, unconditional secure coin flipping is possible to some degree. Although the results of Meyers \cite{8} and Lo, Chau \cite{9} implied the impossibility of perfect quantum CF, there exists quantum CF that can help limit the bias to be strictly less than $\frac{1}{2}$. In retrospect, a lot of progress has been made along the way of exploring the protocols with smaller bias. The first quantum SCF protocol was provided by Aharanov \emph{et al}. \cite{2} with a bias of 0.354\cite{111}. Then Spekkens and Rudolph devised a protocol with a bias of 0.309\cite{111}. Subsequently Ambainis \cite{3} and, independently, Spekkens and Rudolph \cite{4} cut this bound down to 0.25. Later Colbeck \cite{01} proposed a protocol with the same bias 0.25, but it uses a conceptually different approach compared with previous ones. Rather
than being built on bit-commitment, this protocol works by attempting to share entanglement between two parties,
and then exploiting the resulting quantum correlations to implement a coin toss. Furthermore, their protocol requires only qubits for its implementation, whereas bit-commitment based protocols cannot achieve such a bias without using higher dimensional systems. Unfortunately, Ambainis proved that any protocol with a bias of $\epsilon$ must consist of at least $\Omega(\log\log\epsilon^{-1})$ rounds of communication \cite{3}. Then it was proven by Kitaev\cite{10} that any quantum SCF protocols cannot enjoy a bias less than 0.207, which has now been saturated by Chailloux and Kerenidis's protocol \cite{5} based on Mochon's result \cite{7}. With respect to WCF, Spekkens and Rudolph \cite{6} firstly introduced a family of protocols with a bias of 0.207 and Mochon  then pushed this bias down to 0.192 \cite{61} and finally to arbitrary $\epsilon>0$ \cite{7}. In addition, quantum SCF and WCF have also been studied in the multiparty scenario \cite{113}, multioutcome scenario \cite{114,115}, and in both \cite{116,117}.\\
\indent
	In spite of great progress mentioned above, there is a common limit of early results: practical issues were not taken into account. Under imperfect practical conditions such as losses and noise in the quantum channel or in the quantum memory storage, most protocols will totally fail and the bias $\epsilon$ can exactly reach $\frac{1}{2}$ \cite{114}. Therefore, some authors have proposed random bit-string generation instead of single-shot coin flipping\cite{114}. However, this is not interesting from a quantum cryptographic perspective because the same goal can be achieved with purely classical means \cite{119}.\\
\indent
As the most prevalent practical imperfection in the long distance communication, losses were firstly analyzed in devising new practical protocols. In 2008, Berl\'in \emph{et al}. \cite{11}(see also Ref.~\cite{118}) introduced a loss-tolerant SCF protocol with a bias of 0.4. Before long Aharon \emph{et al}. \cite{12} presented a family of loss-tolerant quantum SCF protocols which achieved a smaller bias than Berl\'in \emph{et al}. \cite{11} at a small rate. Recently, Andr\'{e} Chailloux \cite{13} presented an improved loss-tolerant quantum SCF protocol with bias 0.359, by extending Berl\'in \emph{et al}.'s protocol with an encryption step.\\
\indent
	In this article we present a semi-loss-tolerant protocol, which is different from the previous loss-tolerant ones mainly in three aspects:
\begin{enumerate}
\item An EPR pair instead of a qubit is employed to implement the protocol, helping guarantee our protocol semi-loss-tolerant when trying to push the bias down.
\item A quantum memory storage is used and losses in it is taken into account.
\item We find our bias varies with the change of the losses degree of the quantum memory storage, while the bias is independent to the losses in the quantum channel. This proves that there exist CF protocols that are sensitive to losses in the quantum memory storage.
\end{enumerate}
\indent
The rest of the article is organized as follows. In Sec. 2, we give the definitions of loss-tolerance and semi-loss-tolerance. And in Sec. 3, we present our protocol, while details of the two parties' optimal cheating strategies and their maximal biases are obtained in Sec. 4 and 5. In Sec. 6, we obtain the fair scenario by adjusting the value of the free parameter in the protocol. Then some experimental issues concerning the realization of a reliable quantum system implementing our protocol are discussed in Sec. 7. Finally, we make a conclusion and summarize our novelties in Sec. 8.
\section{DEFINITION OF LOSS-TOLERANCE AND SEMI-LOSS-TOLERANCE}
\indent
We say a protocol is loss-tolerant iff it is impervious to any type of losses, including quantum communication channels, measurement devices and quantum memory storage, this definition can be seen in Ref.~\cite{11}. Therefore, protocols in Ref.~\cite{11,12,13} are all loss-tolerant because they are impervious to losses in the quantum channels and measurement devices, the only places that losses may occur in their protocols. \\
\indent
Consider another case, in which the protocol isn't impervious to certain types of losses, but its security varies with the degree of the losses. In other words, the protocol is sensitive to the loss degree of some devices, but with which they are not completely broken like many early protocols. We call this loss-sensitive protocol semi-loss-tolerant. It is clear that the feasibility of this kind of protocols is guaranteed by sufficiently sound environmental factors that the protocols depend on. As discussed below, our EPR-based protocol is semi-loss-tolerant because our bias decreases with the decreasing degree of losses in the quantum memory storage.
\section{EPR-BASED SEMI-LOSS-TOLERANT PROTOCOL}
\indent
As the first one providing a loss-tolerant SCF, Berl\'in \emph{et al}.'s protocol \cite{11} proves that there exist quantum coin flipping protocols outperforming classical ones when taking losses into account. However, it achieves a relatively poor bias compared with previous loss-intolerant protocols. After deliberate consideration, we note that a key factor leading to the relatively high bias in Berl\'in \emph{et al}.'s protocol is that the two parties' bases may be inconsistent. As we know, the basis of measurement is randomly selected by Bob at step 2 in the original protocol(see Ref.~\cite{11}). If the basis is inconsistent with the one that Alice announces, Alice's possible cheating action can't be discovered by Bob, no matter what measurement result Bob gets. We suppose it will cut Alice's bias down if we keep Bob's basis always consistent with Alice's.\\
\indent
To achieve our goal, we let Bob measure the qubit after Alice announces her basis. If Bob detects his qubit and finds nothing wrong with Alice, the outcome of the SCF is successfully generated. If he doesn't detect it, we let the protocol continue to generate the outcome rather than restart, in case of Bob's always successful cheating strategy that he claims detecting nothing if he doesn't like the outcome. Here we present the revised qubit-based protocol:
\begin{enumerate}
\item Alice prepares one state $|\varphi_{a,r_A}\rangle$ from $\{\:|\varphi_{0,0}\rangle=|0\rangle$, $|\varphi_{0,1}\rangle=|1\rangle$, $|\varphi_{1,0}\rangle=\cos\alpha|0\rangle+\sin\alpha|1\rangle$,
$|\varphi_{1,1}\rangle=\sin\alpha|0\rangle-\cos\alpha|1\rangle\:\}$ with basis $a$ and bit $r_A$ chosen independently at random, and then she transmits it to Bob.
\item Bob keeps the received qubit in his quantum memory storage (instead of immediately measures it as described in Ref.~\cite{11}).
\item Bob sends a randomly chosen bit $b$ to Alice.
\item Alice reveals her original $a$ and $r_A$ to Bob.
\item Bob measures the qubit in the quantum memory according to Alice's announcing $a$. If he detects it, whose outcome is denoted as $r_B$, and finds that $r_A\neq r_B$, he aborts the protocol, calling Alice a cheater. If $r_A=r_B$ or even he doesn't detect the qubit, the outcome of the coin flipping is $b\oplus r_A$.
\end{enumerate}
\indent
In this protocol, although the two parties bases are kept consistent, a new problem occurs: if Bob doesn't detect the qubit in his quantum memory, he doesn't know whether the qubit is lost in his quantum memory or in the channel, or even the qubit hasn't been sent by Alice at all. As we know, Bob can't tell whether the qubit is definitely received without measurement. Using this fact, Alice can always succeed in cheating by sending nothing to Bob. To prevent from such attack and keep other properties of our protocol unchanged as much as possible, here we utilize an EPR pair to replace the original qubit. The following is our EPR-based protocol:
\begin{enumerate}
\item Bob prepares a singlet $|\varphi\rangle=\frac{|0_A\rangle |1_B\rangle-|1_A\rangle |0_B\rangle}{\sqrt{2}}$, where the subscripts A and B denote the two entangled particles, then he sends particle A to Alice.
\item	Alice randomly selects a classical bit $a$, where $a=0$ represents that she chooses basis $\{\:|V\rangle=|0\rangle,|V^\bot\rangle=|1\rangle\:\}$ and $a=1$ represents that she chooses basis $\{\:|H\rangle=\cos\alpha|0\rangle+\sin\alpha|1\rangle,|H^\bot\rangle=\sin\alpha|0\rangle-\cos\alpha|1\rangle(0\leq\alpha\leq{\frac{\pi}{2}})\:\}$, then she measures particle A along the basis she chooses. In the following discussions let $|V_A\rangle$ and $|H_A\rangle$ correspond to $r_A=0$, and $|V_A^\bot\rangle$ as well as $|H_A^\bot\rangle$ correspond to $r_A=1$, where $r_A$ denotes the outcome of Alice's measurement. The same principle applies to Bob's later measurement, that is, $|V_B\rangle$ and $|H_B\rangle$ correspond to $r_B=0$, and $|V_B^\bot\rangle$ as well as $|H_B^\bot\rangle$ correspond to $r_B=1$, where $r_B$ denotes the outcome of Bob's measurement.
\item	If Alice successfully detects the particle, she asks Bob to proceed the protocol, otherwise, she asks Bob to restart the protocol.
\item	Bob sends Alice a randomly selected classical bit $b$.
\item	Alice informs Bob of her selected $a$ and outcome $r_A$.
\item	Bob measures particle B along the basis that $a$ represents. If he successfully detects it and finds $r_A=r_B$, he will abort and claim Alice is cheating. In all other cases the outcome of the coin flipping is given by $b\oplus r_A$.
\end{enumerate}
\indent
In this protocol, Alice's measurement on her part of the EPR pair collapses Bob's part to a random state known to herself. In this case, Bob can always make sure that his part, which is equivalent to the qubit sent by Alice in the above qubit-based protocol, is definitely stored in his quantum memory storage, no matter whether it is lost in the quantum memory or not. And Alice's measurement in the EPR-based protocol can be regarded as the process that she randomly sends a qubit to Bob and Bob then detects whether it is successfully received without measurement. As we know, the later action can't be realized and we just utilize the EPR pair to achieve the same goal.\\
\indent
Let's sum up our analysis process. In order to cut down the bias in Berl\'in \emph{et al}.'s protocol\cite{11}, we firstly let Bob measure the qubit after Alice announces her basis, keeping the two parties' bases consistent. After this revision, a new attack occurs: Bob can always succeed in cheating by the strategy that he claims detecting nothing if he doesn't like the outcome. We solve this threat by letting the protocol continue to generate the outcome even if Bob fails to detect the qubit. After this revision, however, another problem occurs: Alice can always succeed in cheating by sending nothing to Bob. Here we utilize an EPR pair to prevent from such attack. Note that we don't specify the value of $\alpha$ but regard it as a free parameter, which is to be adjusted to make the protocol fair. More details of their cheating strategies will be given in the following parts.
\section{ALICE'S MAXIMAL BIAS}
\indent
Since our protocol is symmetrical for the two outcomes, any cheater enjoys the same difficulty in biasing the outcome to 0 or 1.\\
\indent
It seems that Alice can cheat by claiming that she misses the particle when she is unsatisfied with her outcome at step 3. However, since $b$ has not been given at that time and $H(b\oplus r_A|r_A)=1$, it's meaningless to bias $r_A$ to any value. A more general strategy is that before step 4, she can perform a two-outcome positive operator-valued measure(POVM) with elements $E_0^*$ and $E_1^*=I-E_0^*$ on the received particle, trying to collapse the EPR pair to a certain state, without loss of generality, $\beta(\sqrt{E_0^*}\otimes I)|\varphi\rangle$. Here $\beta=\sqrt{\frac{2}{\langle 0|E_0^*|0\rangle+\langle 1|E_0^*|1\rangle}}$ is the normalization factor. And she can also claim that the particle is lost when she is unsatisfied with the outcome. To simplify the notation, let's define $|\varphi^*\rangle=\beta(\sqrt{E_0^*}\otimes I)|\varphi\rangle$, which denotes the combined quantum state after step 3.\\
\indent
At step 5, she can announce a proper $r_A$ after $b$ is given to get the right $r_A\oplus b$ that she wants. Besides, her only worry must be Bob's outcome $r_B$ in the last step. To pass Bob's test, she should bias $r_B$ to the proper value through her optimal measurement and announcement of her selected $a$. \\
\indent
Without loss of generality, assume Alice wants to bias the outcome to 1, then let's discuss two cases. The first case is Bob announces $b=1$, then Alice is clear that she should declare that $r_A=1\oplus b=1\oplus1=0$ and she wants to bais Bob's $r_B$ to $r_A\oplus1=0\oplus1=1$. Therefore Alice should perform another two-outcome POVM measurement with elements $E_b^{r_A,a}$ and $E_b^{r_A,a\oplus1}=I-E_b^{r_A,a}$ just before step 5.
After her measurement, she should inform Bob of $a=a_0$ and $r_A=0$ if her result is associated with $E_1^{0,a_0\oplus1}$, while inform Bob of $a=a_0\oplus1$ and $r_A=0$ if her result is associated with $E_1^{0,a_0\oplus1}$. Here $a_0$ is a Boolean parameter defined by Alice. Without loss of generality, let $a_0=0$, then the probability that Alice's result is associated with $E_1^{0,0}$ equals $P_{E_1^{0,0}}=\langle\varphi^*|E_1^{0,0}\otimes  I|\varphi^*\rangle$, and the composite state after measurement is $\frac{\sqrt{E_1^{0,0}}\otimes I|\varphi^*\rangle}{\sqrt{P_{E_1^{0,0}}}}$. While the probability that Alice's result is associated with $E_1^{0,1}$ equals $P_{E_1^{0,1}}=\langle\varphi^*|E_1^{0,1}\otimes I|\varphi^*\rangle$, and the composite state after measurement is $\frac{\sqrt{E_1^{0,1}}\otimes I|\varphi^*\rangle}{\sqrt{P_{E_1^{0,1}}}}$. If Bob successfully detects his particle, the probability that he gets $r_B=0$, which implies that Alice succeeds in cheating, after his measurement according to Alice's claiming $a$ is given by
\begin{eqnarray}
P_{S1}&=&P_{E_1^{0,0}}\frac{\langle\varphi^*|\sqrt{E_1^{0,0}}\otimes I}{\sqrt{P_{E_1^{0,0}}}}(I\otimes |V_B\rangle\langle V_B | ) \frac{\sqrt{E_1^{0,0}}\otimes I|\varphi^*\rangle}{\sqrt{P_{E_1^{0,0}}}}+\\\nonumber
&&\:\:\:\:P_{E_1^{0,1}}\frac{\langle\varphi^*|\sqrt{E_1^{0,1}}\otimes I}{\sqrt{P_{E_1^{0,1}}}}(I\otimes |H_B\rangle\langle H_B|) \frac{\sqrt{E_1^{0,1}}\otimes I|\varphi^*\rangle}{\sqrt{P_{E_1^{0,1}}}}\\\nonumber
&=&\langle\varphi^*|(E_1^{0,0}\otimes |V_B\rangle\langle V_B|) |\varphi^*\rangle+\langle\varphi^*|(E_1^{0,1}\otimes |H_B\rangle\langle H_B|) |\varphi^*\rangle\\\nonumber
\end{eqnarray}
Insert $|\varphi^*\rangle=\sqrt{E_0^*}\otimes I|\varphi\rangle$ and $E_1^{0,1}=I-E_1^{0,0}$ to above expression, we get
\begin{eqnarray}
P_{S1}&=&\frac{\langle V_A^\bot|\sqrt{E_0^*}E_1^{0,0}\sqrt{E_0^*}|V_A^\bot\rangle-\langle H_A^\bot|\sqrt{E_0^*}E_1^{0,0}\sqrt{E_0^*}|H_A^\bot\rangle+\langle H_A^\bot|E_0^*|H_A^\bot\rangle}{\langle 0|E_0^*|0\rangle+\langle 1|E_0^*|1\rangle}
\end{eqnarray}

In the other case, Bob announces $b=0$, let's assume Alice performs POVM measurement with elements $E_b^{r_A,a}=E_0^{0,0}$ and  $E_b^{r_A,a\oplus1}=E_0^{0,1}=I-E_0^{0,0}$ just before step 5. Similarly, we can obtain that the probability that Alice succeeds in cheating is given by
\begin{eqnarray}
P_{S0}&=&\frac{\langle V_A|\sqrt{E_0^*}E_0^{0,0}\sqrt{E_0^*}|V_A\rangle-\langle H_A|\sqrt{E_0^*}E_0^{0,0}\sqrt{E_0^*}|H_A\rangle+\langle H_A|E_0^*|H_A\rangle}{\langle 0|E_0^*|0\rangle+\langle 1|E_0^*|1\rangle}
\end{eqnarray}

Note the two cases (i.e. Bob announces $b=0$ and $b=1$) happens with the same probability $\frac{1}{2}$, combining them together, we obtain the probability that Alice succeeds in cheating is given by
\begin{eqnarray}
P_{S}&=&\frac{P_{S0}+P_{S1}}{2}
\end{eqnarray}

Clearly, Alice wants to maximize $P_S$ in order to increase her probability to succeed. Using numerical method, we firstly fix $\alpha$ then randomly select $10^6$ qualified $(E_0^*, E_0^{0,0}, E_1^{0,0})$s to test the value of $P_S$. Then we find $P_S$ will get its maximum value if $(E_0^*, E_0^{0,0}, E_1^{0,0})=(I, |\lambda_0^{max}\rangle\langle\lambda_0^{max}|, |\lambda_1^{max}\rangle\langle\lambda_1^{max}|)$ with $|\lambda_0^{max}\rangle(|\lambda_1^{max}\rangle)$ being the maximum eigenvalue of the operator $|V_A\rangle\langle V_A|-|H_A\rangle\langle H_A|(|V_A^\bot\rangle\langle V_A^\bot|-|H_A^\bot\rangle\langle H_A^\bot|)$ and $\lambda_0^{max}(\lambda_1^{max})$ is the corresponding eigenstate.\\

\indent
After simple calculation, we find $\lambda_0^{max} = \lambda_1^{max} = \sin\alpha$, together with $(E_0^*, E_0^{0,0}, E_1^{0,0})=(I, |\lambda_0^{max}\rangle\langle\lambda_0^{max}|, |\lambda_1^{max}\rangle\langle\lambda_1^{max}|)$, we obtain
\begin{eqnarray}
P_{S}&=&\frac{P_{S0}+P_{S1}}{2}\\\nonumber
&\leq&\frac{\frac{\sin\alpha+1}{2}+\frac{\sin\alpha+1}{2}}{2}\\\nonumber
&=&\frac{\sin\alpha+1}{2}
\end{eqnarray}
So the maximal probability that Alice passes Bob's test if Bob successfully detects his particle equals $P_{S}^{max}=\frac{\sin\alpha+1}{2}$. It remains to analytically prove that the maximum value of $P_{S}$ is $\frac{\sin\alpha+1}{2}$ and it remains to figure out whether $(E_0^*, E_0^{0,0}, E_1^{0,0})=(I, |\lambda_0^{max}\rangle\langle\lambda_0^{max}|, |\lambda_1^{max}\rangle\langle\lambda_1^{max}|)$ is the only optimal solution.\\
\indent
However, if Bob doesn't detect his check particle, our protocol still proceeds, and in such a case, Bob misses the chance to detect Alice's possible cheating and a dishonest Alice will definitely succeed. To combine the two cases together, let $p(0\leq p\leq1)$ denote the probability that Bob fails to detect particle B, then the maximum probability that Alice succeeds is given by
\begin{eqnarray}
P_A^{max}&=&(1-p)\times P_{S}^{max}+p\times 1\\\nonumber
&=&(1-p)\times \frac{\sin\alpha+1}{2}+p\times 1\\\nonumber
&=&\frac{1+p+(1-p)\sin\alpha}{2}
\end{eqnarray}
Thus the maximum bias for Alice equals
\begin{eqnarray}
\epsilon_A^{max}=\frac{p+(1-p)\sin\alpha}{2}
\end{eqnarray}
\indent
Comparing Alice's bias in our protocol with that in Ref.~\cite{11}, let $\epsilon_A^M$ and $\epsilon_A^B$ respectively denote Alice's bias in our protocol and that in \cite{11}, we have
\begin{eqnarray}
\Delta\epsilon_A&=&\epsilon_A^M-\epsilon_A^B\\\nonumber
&=&\frac{p+(1-p)\sin\alpha}{2}-\frac{1+\sin\alpha}{4}\\\nonumber
&=&\frac{(2p-1)(1-\sin\alpha)}{4}
\end{eqnarray}
If $p<\frac{1}{2}$, that means the quantum memory is good enough, then $\Delta\epsilon_A<0$, which means Alice's bias is successfully cut down by our revisions.
\section{BOB'S MAXIMAL BIAS}
\indent
Assume Bob wants to bias the outcome to 0, the most general cheating strategy for him is to firstly choose $b$ to be sent at step 4 and then prepare an entangled state $|\varphi'\rangle$ instead of $|\varphi\rangle$ so that the probability that Alice's $r_A$ equals $b\oplus0$ will reach the maximum.\\
\indent
	Without loss of generality, assume Bob selects $b=0$ and prepares $|\varphi'\rangle=\sqrt{\lambda}|H_A'\rangle |V_B' \rangle+\sqrt{1-\lambda}| {H_A'}^\bot \rangle |{V_B'}^\bot \rangle (\frac{1}{2}\leq\lambda\leq1)$, where $|\varphi'\rangle$ is written in its Schmidt decomposition form and
$|H_A'\rangle=\cos\beta|0\rangle+\sin\beta|1\rangle,|{H_A'}^\bot\rangle=\sin\beta|0\rangle-\cos\beta|1\rangle
(0\leq\beta\leq\frac{\pi}{2})$,while $|V_B' \rangle$ and $|{V_B'}^\bot \rangle$ represent some choice of orthogonal
states on Bob's space. Thus the state of particle A can be written as $\rho_A=\lambda|H_A'\rangle\langle H_A'|+(1-\lambda)|{H_A'}^\bot\rangle\langle{H_A'}^\bot|$. Since Alice is honest, she will carry out the measurement on the particle according to $a$ she randomly selects. Then the probability that she gets $r_A=0\oplus0=0$, which implies that Bob succeeds in cheating, is given by
\begin{eqnarray}
P_B&=&\frac{1}{2}\langle V_A|\rho_A|V_A\rangle+\frac{1}{2}\langle H_A|\rho_A|H_A\rangle\\\nonumber
&=&\frac{1}{2}+\frac{(2\lambda-1)\cos\alpha\cos{(\alpha-2\beta)}}{2}\\\nonumber
&\leq&\frac{1}{2}+\frac{\cos\alpha}{2}
\end{eqnarray}
\indent This bound can be saturated iff $\lambda=1$ and $\beta= \frac{\alpha}{2}$. In this case $|\varphi'\rangle=|H_A'\rangle |V_B' \rangle$, which is a product state so that $|V_B'\rangle$ is unnecessary for Bob. In conclusion, the best cheating strategy for Bob is to send $|H_A'\rangle=\cos{\frac{\alpha}{2}}|0\rangle+\sin{\frac{\alpha}{2}}|1\rangle$ to Alice at step 1 and then claims $b=0$ at step 3, then the probability that he succeeds in biasing the outcome to 0 equals $P_B^{max}=\frac{1}{2}+\frac{\cos\alpha}{2}$, and the maximal bias for Bob equals
\begin{eqnarray}
\epsilon_B^{max}=\frac{\cos\alpha}{2}.
\end{eqnarray}
\indent
We notice that by this strategy, Bob losses the chance to check Alice's cheating because he has no check particle to be entangled with the particle sent to Alice. However, this is not a flaw for Bob in our protocol, since we don't consider the situation when the two parties are simultaneously dishonest. Also note that Bob's bias is equal to that in Berl\'in $et$ $al$.'s protocol\cite{11}, which means our revisions have no effects on Bob's  bias.
\section{FAIR SCENARIO}
\begin{figure} [htbp]
\vspace*{13pt}
\centerline{\epsfig{file=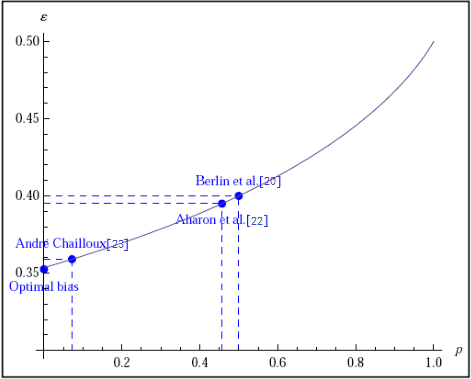, width=8.2cm}} 
\vspace*{13pt}
\fcaption{\label{123}Maximal fair bias is a function of $p$, it decreases with decreasing $p$, and our optimal bias outperforms all the previous ones.}
\end{figure}

\indent
To make our protocol fair we must adjust the free parameter $\alpha$ so that
\begin{eqnarray}
\epsilon_A^{max}=\epsilon_B^{max}
\end{eqnarray}
Inserting Eq.(7) and Eq.(10) into Eq.(11), we must have
\begin{eqnarray}
\frac{p+(1-p)\sin\alpha}{2}=\frac{\cos\alpha}{2}
\end{eqnarray}
Solving for $\alpha$ in terms of $p$ we get
\begin{eqnarray}
\alpha=\arcsin \frac{p^2-p+\sqrt{2-2p}}{p^2-2p+2}
\end{eqnarray}
By implying Eq.(13) to Eq.(7) and Eq.(10), we get the bias in the fair scenario, which is given by
\begin{eqnarray}
\epsilon(p)=\epsilon_A^{max}=\epsilon_B^{max}=\frac{p+(1-p)\sqrt{2-2p}}{2(p^2-2p+2)}
\end{eqnarray}
\indent
We find that the maximal fair basis $\epsilon(p)$ monotonously decreases as $p$ decreases(see Fig.~\ref{123}), which means the more likely Bob successfully detects his particle, the securer the protocol can be. From the definition given in Sec. 2, we can call our protocol semi-loss-tolerant. This characteristic is totally different from all the loss-tolerant protocols whose biases are independent of the degree of the losses.

\indent Comparing our bias with the previous loss-tolerant ones, we notice that $\epsilon(0.5)=0.4$, $\epsilon(0.457)=0.395$ and $\epsilon(0.072)=0.359$, which respectively represents biases in Refs. \cite{11,12,13}. In particular, $\epsilon(0)=0.3536$, which means if the quantum memory storage approaches perfect, we can achieve a lower bias, 0.3536(see Fig.~\ref{123}). Also note that even if $p=0.5$, that means the quantum memory is not very reliable, our protocol can still achieve the bias $0.4$.
\section{EXPERIMENTAL ISSUES}
\indent
We have to admit that the improved performance relative to previous protocols only occurs when the degree of losses in the quantum memory is small enough. For example, to be better than bias 0.359 in Ref.\cite{13}, the loss rate $p$ of the quantum memory in our protocol must be less than $7.2\%$. Fortunately, entangled trapped atom-photon systems display the required behavior. It has been experimentally demonstrated by Blinov et al.\cite{02} that using entangled trapped atom-photon systems can help us realize the process of EPR generation, transmission and storage that are required in our protocol. Specifically, the transmitted particle in our protocol can be realized by the polarization of the emitted photon, while another one stored in the quantum memory can be represented by the internal atomic qubit levels, stored in $^2{S_{\frac{1}{2}}}$ hyperfine ground states. Experiment results has shown that the success probability of detecting the transmitted photon is $P\approx1.6\times10^{-4}$. The experiment repetition rate is $R <2\times10^3 s^{-1}$, resulting in an entanglement generation rate $R_1=PR < 0.3 s^{-1}$. That means in our protocol, Alice is expected to detect the particle within $\frac{1}{R_1^{max}}\approx3.3s$. What's more, the coherence time of trapped ion is very long and the loss of ion quantum memory is negligible for practical applications. That means the sufficiently small loss rate of the quantum memory required in our protocol can be achieved. Taking the efficiency with which the state of the quantum memory can be read out, which is almost $100\%$, into consideration, we obtain the total efficiency of our protocol is almost $R_1^{max}\times 100\% = 0.3 s^{-1}$.\\
\indent
Although not the main concern of this paper, the fidelity of such a memory will inevitably influence the performance of our protocol. As we know, an imperfect fidelity will create errors, which will decrease the storage time. Hence, the longer is the distance between Alice and Bob, the longer will the required storage time be, and the higher will the error rate be.\\
\indent
Considering the motivation behind the introduction of loss-tolerant protocols is experimental, our scenario has been theoretically demonstrated experimentally feasible. Recently, Berl\'in \emph{et al}. Ref.\cite{888} implemented the first experimental demonstration of a loss-tolerant quantum coin flipping protocol using single  qubits, and  we are interested to see practical realizations of such a protocol with EPR pairs.
\section{CONCLUSION}
\indent
We have presented an EPR-based semi-loss-tolerant SCF protocol, and the novelties in our manuscript can be summarized as:
\begin{enumerate}
\item We have proposed a novel approach to solve the issue implied in Ref.~\cite{11}, i.e., the two parties' bases may be inconsistent, which we think is a key factor that leads to the relatively higher bias in Ref.~\cite{11}.
\item We introduce the EPR pair to prevent Alice's sending-nothing cheating strategy.
\item As one of the most important indicator in a CF, the bias in our protocol performs better than the previous loss-tolerant ones in the presence of sufficiently small losses in the quantum memory storage.
\item We find our bias varies with the change of the losses degree of the quantum memory storage, while the bias is independent to the losses in the quantum channel just like the previous loss-tolerant CF protocols. This proves that there exist semi-loss-tolerant CF protocols that are sensitive to losses in the quantum memory storage.
\item This is the first manuscript taking the losses in the quantum memory into account. As discussed in Sec VII of Ref.~\cite{12}, when introducing a loss-tolerant WCF protocol, ``it would seem that a major difficulty is that at the end of a WCF protocol the losing party usually verifies the outcome by measuring a quantum system that has been kept in a quantum memory storage. Hence, in this scenario the losing party can always avoid losing by claiming to have lost the stored system". We suppose that step 6 in our EPR-based protocol provides significant inspiration to this problem.
\item We discussed  that since trapped ions' state can be read-out with almost 100\% efficiency, implementing our protocol with a trapped ion entangled with a single photon makes our proposal potentially feasible in practice. Also, we demonstrated that losses in the transmission channel only affect the efficiency instead of the bias of our protocol.
\end{enumerate}
\indent
Moreover, there is still a problem when implementing this protocol. Before starting the protocol, Alice and Bob must negotiate to choose a proper $\alpha$ according to $p$ and Eq.(13). The question is how to obtain the real $p$, which is a parameter of Bob's machine. If $p$ is claimed by Bob himself, he can cheat by claiming a relatively larger value than the real $p_0$. In such a case, Alice's bias $\epsilon_B$ equals $\epsilon(p_0)$, while Bob's bias $\epsilon_A$ is actually larger than $\epsilon(p_0)$. That means the protocol is actually unfair. A possible way to solve this problem is that Alice sets a threshold for $p$, if Bob's announcing $p$ surpasses the threshold, she will refuse to implement the protocol with Bob.

\nonumsection{Acknowledgements}
\noindent
This work is supported by NSFC (Grant Nos. 61170270, 61100203, 60903152, 61003286, 61121061), NCET (Grant No. NCET-10-0260), Beijing Natural Science Foundation(Study on the divice-independent quantum key distribution protocol and its security, Grant No. 4112040),SRFDP (Grant No. 20090005110010),The National Basic Research Program of China (Grant No. 2010CB923200).

\nonumsection{References}


\noindent
\begin{thebibliography}{99}
\bibitem{1}M. Blum (1981), {\it Coin Flipping by Telephone}, Advances in Cryptology: A Report on CRYPTO'81 (Santa-Barbara, CA), pp. 11.
\bibitem{2}D. Aharonov, A. Ta-Shma, U. Vazirani, and A. C. Yao (2000), {\it Quantum Bit Escrow}, Proceedings of the 32nd Annual Symposium on Theory of Computing (New York), pp. 705.
\bibitem{3}A. Ambainis (2001), {\it A new protocol and lower bounds for quantum coin flipping}, Proceedings of the 33rd Annual Symposium on Theory of Computing (New York), pp. 134.
\bibitem{4}R.W. Spekkens and T. Rudolph (2001), {\it Degrees of concealment and bindingness in quantum bit commitment protocols}, Phys. Rev. A, Vol. 65, pp. 012310.
\bibitem{01}Roger Colbeck (2007), {\it An Entanglement-Based Protocol For Strong Coin Tossing With Bias 1/4}, Phys. Lett. A, Vol. 362, pp. 390-392.
\bibitem{5}A. Chailloux and I. Kerenidis (2009), {\it Optimal quantum strong coin flipping}, Proceedings of the 50th Annual IEEE Symposium on the Foundations of Computer Science (Atlanta, GA), pp. 527.
\bibitem{61}C. Mochon (2004), {\it Quantum weak coin-flipping with bias of 0.192}, quant-ph/0403193.
\bibitem{6}R. W. Spekkens and T. Rudolph (2002), {\it Quantum Protocol for Cheat-Sensitive Weak Coin Flipping}, Phys. Rev. Lett., Vol. 89, pp. 227901.
\bibitem{7}C. Mochon (2007), {\it Quantum weak coin flipping with arbitrarily small bias}, quant-ph/0711.4114.
\bibitem{8}Dominic Mayers (1997), {\it Unconditionally secure quantum bit commitment is impossible}, Phys. Rev. Lett., Vol. 78, pp. 3414-3417.
\bibitem{9}Hoi-Kwong Lo and H. F. Chau (1997), {\it Is quantum bit commitment really possible?}, Phys. Rev. Lett., Vol. 78, pp. 3410-3413.
\bibitem{111} R. W. Spekkens and T. Rudolph (2002), {\it Optimization of coherent attacks in generalizations of the BB84 quantum bit commitment protocol}, Quantum Inf. Comput., Vol. 2, pp. 66.
\bibitem{10}A. Kitaev (unpublished). The proof is reproduced in Ref.~\cite{113}.
\bibitem{113}A. Ambainis, H. Buhrman, Y. Dodis, and H. R\"{o}hrig (2004), {\it Multiparty quantum coin flipping}, Proceedings of the 19th IEEE Annual Conference on Computational Complexity (Amherst, MA), pp. 250.
\bibitem{114}J. Barrett and S. Massar (2004), {\it Quantum coin tossing and bit-string generation in the presence of noise}, Phys. Rev. A, Vol. 69, pp. 022322.
\bibitem{115}J. Barrett and S. Massar (2004), {\it Security of quantum bit-string generation}, Phys. Rev. A, Vol. 70, pp. 052310.
\bibitem{116}N. Aharon and J. Silman (2010), {\it Quantum dice rolling: a multi-outcome generalization of quantum coin flipping}, New J. Phys., Vol. 12, pp. 033027.
\bibitem{117}M. Ganz (2009), {\it Quantum Leader Election}, quant-ph/0910.4952.
\bibitem{119}H. Buhrman, M. Christandl, M. Koucky, Z. Lotker, B. Patt-Shamir, and N. K. Vereshchagin (2007), {\it High Entropy Random Selection Protocols}, Approximation, Randomization, and Combinatorial Optimization. Algorithms and Techniques, Lecture Notes in Computer Science (Springer, Berl\'in), Vol. 4627, pp. 366¨C379.
\bibitem{11}G. Berl\'in, G. Brassard, F. Bussi\`{e}res, and N. Godbout (2009), {\it Fair loss-tolerant quantum coin flipping}, Phys. Rev. A, Vol. 80, pp. 062321.
\bibitem{118}A. T. Nguyen, J. Frison, K. Phan Huy, and S. Massar (2008), {\it Experimental quantum tossing of a single coin}, New J. Phys., Vol. 10, pp. 083037.
\bibitem{12}N. Aharon, S. Massar, and J. Silman (2010), {\it A family of loss-tolerant quantum coin flipping protocols}, Phys. Rev. A, Vol. 82, pp. 052307.
\bibitem{13}A. Chailloux (2011), {\it Improved Loss-Tolerant Quantum Coin Flipping}, quant-ph/1009.0044.
\bibitem{02}B. B. Blinov, D. L. Moehring, L.- M. Duan \& C. Monroe (2004), {\it Observation of entanglement between a single trapped atom
and a single photon}, Nature, Vol. 428, pp. 153.
\bibitem{888}Guido Berl\'in, Gilles Brassard, F\'elix Bussi\'eres, Nicolas Godbout, Joshua A. Slater \& Wolfgang Tittel (2011), {\it Experimental loss-tolerant quantum coin flipping}, Nature Communications 2, Article number: 561 doi:10.1038/ncomms1572.
\end{thebibliography}
\end{document}